\documentclass[12pt]{article}
\usepackage[english,russian]{babel}
\usepackage{amsmath}
\usepackage{amsfonts}
\usepackage{amscd}
\usepackage{amsthm}
\usepackage{amssymb}
\usepackage{latexsym}
\newcommand{\beq}{\begin{equation}}
\newcommand{\eeq}{\end{equation}}
\newcommand{\bit}{\begin{itemize}}
\newcommand{\eit}{\end{itemize}}
\newcommand{\rd}{\rm{d}}
\newcommand{\ri}{\rm{i}}
\begin{document}

{\large\bf M. A. Sokolov}
\vspace{0.5cm}

\centerline{\Large\bf Generating functions of Chebyshev polynomials}
\vspace{0.5cm}
\centerline{\Large\bf in three variables \footnote{
This work has been supported by RFFR grant no. 15-01-03148-а}}

\begin{flushright}
{\it To Petr Petrovich Kulish} \\
{\it on occasion of his 70th birthday}
\end{flushright}

{\bf 1.} The aim of the present work is to obtain generating functions of 
Chebyshev polynomials in three variables.
Chebyshev polynomials in several variables associated with the root systems 
of the simple Lie algebras studied intensively
in the last few decades \cite{K1} - \cite{LU}. Applications can be found for them in different areas 
of mathematics, as well as in the physical investigations. The examples of such applications with 
the references therein may be found in works \cite{RM-K} - \cite{G}.

Multivariate Chebyshev polynomials are natural generalization of the classical ones. 
The classical Chebyshev polynomials of the first kind $T_n(x)$ are defined by the 
following formula (see for example \cite{S})

\begin{equation}
\label{0-1}
T_n(x)=T_n(\cos{\phi})=\cos{n\phi},\quad n=0,1,..,
\end{equation}
and they satisfy the following three-term recurrence relation

\begin{equation}
\label{0-2}
T_{n+1}(x)=2xT_n(x)-T_{n-1}(x),\quad n\ge 1.\nonumber
\end{equation}
The classical Chebyshev polynomials of the second kind $U_n(x)$ are defined by the formula

\begin{equation}
\label{0-3}
U_n(x) = \frac{\sin{(n+1)\phi}}{\sin{\phi}},\quad n\ge 0, \quad  x = \cos{\phi},
\end{equation}
which is different from (\ref{0-1}), but they satisfy the same recurrence relation (\ref{0-2}).
The initial conditions for the above polynomials

\begin{equation}
\label{0-4}
T_0(x) = 1, \quad  T_1(x) = x,\,\,\,  U_0(x) =1,\quad U_1(x)=2x
\end{equation}
together with the recurrence relation (\ref{0-2}) determine them uniquely without 
reference to (\ref{0-1}), (\ref{0-3}).

It is known that the function $\cos{n\phi}$ can be treated as an invariant mean of 
the exponential function on the Weyl group of the $A_1$ algebra root system

\begin{equation}
\label{0-5}
\cos{n\phi} = \frac1{2}(e^{\ri n\phi}+e^{-\ri n\phi}).
\end{equation}
Transition to a function which is invariant mean on a Weyl group $W$ of other root 
system leads us to generalization of the classical Chebyshev polynomials to polynomials 
in several variables \cite{K1} - \cite{B}.
Let us remind shortly the way in which such $W$-invariant functions can be constructed.

Let $R(L)$ be a reduced root system of any simple Lie algebra $L$. Such a system is a 
set of vectors in
$d$-dimensional Euclidean space $E^d$ supplied with a scalar product $(.,.)$. 
The system $R$ is uniqually determined by the base of simple roots 
$\alpha_i,\,i=1,..,d$ and a finite group $W(R)$, which is generated by the base reflections. 
This group is called a Weyl group.

Generating elements $w_i,\,i=1,..,d$ of the Weyl group act on the simple roots according 
to $w_i\alpha_i =-\alpha_i.$ The system $R$ is closed under the action of a Weyl group. 
The elements $w$ of $W(R)$ acts on any vector $x\in E^d$ by the rule

\begin{equation}
\label{0-6}
w_ix=x-\frac{2(x,\alpha_i)}{(\alpha_i,\alpha_i)}\alpha_i.
\end{equation}
To each root $\alpha\in R$ corresponds a coroot  defined by the formula

$$\alpha^{\vee}=\frac{2\alpha}{(\alpha,\alpha)}.$$
The base of fundamental weights $\lambda_i,\,i=1,..,d$ is dual to the coroot 
base $\alpha^{\vee}_i,\,i=1,..,d$

\begin{equation}
\label{dop-1}
(\lambda_i,\alpha^{\vee}_j)=\delta_{ij}
\end{equation}
(we identify the dual space ${E^d}^*$ with $E^d$). It is worth noting that the both 
bases are not orthogonal.

Multivariate Chebyshev polynomials of the first kind can be defined by the following function

\begin{equation}
\label{0-7}
\Phi_{\bf n}({\boldsymbol{\phi}}) =
\sum\limits_{w\in {\mbox{\footnotesize W}}}e^{2\pi\ri (\emph{w}\,{\bf n},{\boldsymbol{\phi}})},
\end{equation}
where the vector $\bf n \in E^d$ is given in the fundamental weights base and the vector 
${\boldsymbol{\phi}}$ is given in the coroot base $\{\alpha^\vee_i\}$
\begin{equation}
\label{0-8}
{\bf n}=\sum_{i=1}^d\,n_i\lambda_i \quad n_i\in Z, \quad  {\boldsymbol{\phi}}=
\sum_{i=1}^d\,\phi_i\alpha_i^\vee \quad \phi_i\in [0,1), \quad (\lambda_i,\alpha_j^\vee)=\delta_{ij}.
\end{equation}
Evidently the function $\Phi_{\bf n}({\boldsymbol{\phi}})$ is invariant with respect to 
action of the corresponding Weyl group,

$$\Phi_{\tilde w\bf n}({\boldsymbol{\phi}})=\Phi_{\bf n}({\boldsymbol{\phi}}), \,\tilde w\,\in W.$$
Many of the useful properties of $\Phi_{\bf n}({\boldsymbol{\phi}})$ were discussed in the review \cite{KP}, 
where these functions were called orbit functions.

In the case of the Lie algebra $A_1$ there is the only simple root $\alpha = \alpha^\vee$ 
(we use the standard normalization $(\alpha,\alpha)=2$) and  there is the only fundamental 
weight $\lambda = \alpha/2$.
The Weyl group $W(A_1)$ consist of the identity $e$ and the idempotent $w:\,w\alpha = -\alpha$. 
Thus, the orbit function (\ref{0-7}) has the form

$$
\Phi_{\bf n}({\boldsymbol{\phi}}) =e^{2\pi\ri ({\bf n},{\boldsymbol{\phi}})} +
e^{2\pi\ri (\emph{w}\,{\bf n},{\boldsymbol{\phi}})} = e^{\pi\ri n\phi} + e^{-\pi\ri n\phi},
$$
which is identical with (\ref{0-5}) up to the factor $1/2$ and redefinition of phase $\phi$.

Multivariate Chebyshev polynomials of the second kind can be defined by the Weyl character formula

\begin{equation}
\label{0-9}
U_{\bf n}({\boldsymbol{\phi}})=\frac{\sum\limits_{w\in {\mbox{\footnotesize W}}}\det{w}\,\,
e^{2\pi\ri (\emph{w}\,({\bf n}+\boldsymbol{\rho}),{\boldsymbol{\phi}})}}{\sum\limits_{w\in 
{\mbox{\footnotesize W}}}\det{w}\,\,e^{2\pi\ri (\emph{w}\,{\boldsymbol{\rho}},{\boldsymbol{\phi}})}}
=\frac{\Phi_{\bf n+\boldsymbol{\rho}}^{as}}{\Phi_{\boldsymbol{\rho}}^{as}},
\end{equation}
where $\det{w} = (-1)^{\ell (\emph{w})}$  and ${\ell}(\emph{w})$ is the minimal number of Weyl 
group generating elements $w_i$ required for expressing $w$  as a product of  $w_i$. 
$\boldsymbol{\rho}$ is the Weyl vector which is equal to the sum of the fundamental weights. 
It is essential for our purpose that the numerator and denominator in (\ref{0-9}) 
can be represented in the following  form

\begin{equation}
\label{dop-2}
\Phi_{\bf m}^{as} = \Phi_{\bf m}^{as+}-\Phi_{\bf m}^{as-}= \sum\limits_{w\in {\mbox{\footnotesize W}},\,
\det{w}=1}\,e^{2\pi\ri (\emph{w}\,{\bf m},{\boldsymbol{\phi}})}-\sum\limits_{w\in 
{\mbox{\footnotesize W}},\,\det{w}=-1}\,e^{2\pi\ri (\emph{w}\,{\bf m},{\boldsymbol{\phi}})},
\end{equation}
where $\bf m$ is any vector written  in the fundamental weight base.

{\bf 2.} Generating functions are a powerful tool as in the theory of classical orthogonal 
polynomials, as in various applications. It is evidently that generating functions are important 
for the studying in multivariate Chebyshev polynomials too. In the work \cite{DL} the 
generating functions of Chebyshev polynomials of both kinds associated with the Lie algebra $A_2$ 
were obtained in explicit form. Recently there was proposed a method of obtaining generating function 
of  multi\-va\-riate Chebyshev polynomials of both kinds \cite{DKS}. This method uses only an 
explicit form of  $W$ - invariant functions (\ref{0-7}) and is independent from knowledge of 
recurrent relations.  In \cite{DKS} the generating functions of  Chebyshev polynomials in 
two variables associated with the Lie algebras $C_2$ and $G_2$  were obtained by this method. 
The method can be summarized as follows.

Since the components of ${\bf n}$ are integer, scalar product in the function $\Phi_{\bf n}$ 
(\ref{0-7}) can be represented in the form 
$(\emph{w}\,{\bf n},{\boldsymbol{\phi}}) = \sum_k(w\lambda_k,{\boldsymbol{\phi}})n_k$, 
and the function itself can be written as

\begin{equation}
\label{dop-6}
\Phi_{\bf n} = \sum\limits_{w\in {\mbox{\footnotesize W}}}\prod_k\left(e^{2\pi\ri 
(w\lambda_k,{\boldsymbol{\phi}})}\right)^{n_k} = {\rm tr}\left(\prod_kM_k^{n_k}\right).
\end{equation}
In this formula $M_k$ are the following diagonal matrices
$$M_k={\rm diag}
(e^{2\pi\ri (w_1\lambda_k,{\boldsymbol{\phi}})},
e^{2\pi\ri (w_2\lambda_k,{\boldsymbol{\phi}})},....,
e^{2\pi\ri (w_{|W|}\lambda_k,{\boldsymbol{\phi}})}),$$
$w_i$ are elements of the Weyl group, and   $|W|$  is the number of elements 
in the Weyl group $W$.

Define the matrices $R_k=(I_{|W|}-p_kM_k)^{-1}$, where  $I_{|W|}$ is the identity 
matrix of size $|W|$ and $p_k$ are real parameter. In these notations $\Phi_{\bf n}$ has the form

$$
\Phi_{{\bf n}}({\boldsymbol{\phi}})=\Phi_{{n_1,..,n_d}}({\boldsymbol{\phi}}) =
\frac1{n_1!..n_d!}\left.\frac{\rd^{n}}{\rd^{n_1}p_1..\rd^{n_d}p_n}
\left({\rm tr}(R_{p_1}..R_{p_d})
\right)\right|_{p_1= .. =p_d = 0}.
$$
Let us introduce the new independent variables defined by
$$
x_i\,=\,\Phi_{{\bf e}_i}({\boldsymbol{\phi}}),\quad {\bf e}_i = 
(\overbrace{0,..,0}^{i-1},1,\overbrace{0,..,0}^{d-i}).
$$
The simple structure of the matrices $R_{p_k}$ allows us to express the coefficients
$\Phi_{{n_1,..,n_d}}({\boldsymbol{\phi}})$ of the function

$$
F_{p_1,..,p_d}^I =
{\rm tr}(R_{p_1}..R_{p_d}) = \sum\limits_{n_1..n_d\ge 0}\Phi_{{n_1,..,n_d}}
({\boldsymbol{\phi}})p_1^{n_1}..p_d^{n_d}
$$
in the terms of $x_i$.  It is evidently, that the function $F_{p_1,..,p_d}^I$ can  
be considered as a generating function of the multivariate Chebyshev polynomials of the first kind.

For calculation of generating functions of multivariate Chebyshev polynomials of the 
second kind is required only a minor modification of the above method. The modification 
consists in representation of the functions with alternating signs from the Weyl character 
formula (\ref{0-9}) as a difference of two terms (\ref{dop-2}). This representation allows 
us to use the calculation scheme presented in this section.

{\bf 3.} The Chebyshev polynomials in three variables were studied in the work \cite{SJ} 
out of the Lie algebraical context. In particular, in this work were obtained the 
following recurrence relations

\begin{equation}
\begin{split}
T_{k+1,m,n}&=4zT_{k,m,n}-T_{k-1,m,n+1}-T_{k,m+1,n-1}-T_{k,m-1,n}, {} \nonumber\\
T_{k,m+1,n} &= 6rT_{k,m,n}-T_{k+1,m-1,n+1}-T_{k-1,m,n+1}-T_{k+1,m,n-1}-T_{k-1,m+1,n-1}-
T_{k,m-1,n}.{}\label{ch-01}\\
T_{k,m,n+1} &= 4\bar z T_{k,m,n} - T_{k,m+1,n-1} - T_{k+1,m-1,n} - T_{k-1,m,n}.{} \nonumber
\end{split}
\end{equation}
and  were calculated some first polynomials using these relations and 
the natural initial conditions  
$T_{0,0,0}=1$, $T_{1,0,0}=z$, $T_{0,1,0}=r$, $T_{0,0,1}=\bar z$.

In what follows we use the representation of the roots of the Lie algebra $A_3$ 
by the vectors of the Euclidean space $E^4$  with the standard scalar product 
$(.,.)_E$ (see for example \cite{NBurb}). In the formulas below the index $E$ will be dropped.
Let $e_i,\,i=1,2,3,4$ be a system of orthogonal normalized vectors 
$(e_i,e_j)=\delta_{ij}$ of $E^4$.
In this case the triple of the fundamental roots of $A_3$ is defined as 
$\alpha_i = e_i - e_{i+1},\,i=1,2,3$. The positive roots consist the set
$\alpha_{ij}= e_i - e_{j},\,1\le i < j \le 4$. All the roots have the norm $|\alpha_i|=\sqrt{2}$.

In this representation  the fundamental weights are defined by the formulas

\begin{equation}\label{dop-4}
\begin{split}
\lambda_1&=\frac{3}{4}\alpha_1+\frac{1}{2}\alpha_2+\frac{1}{4}\alpha_3=
\frac{3}{4}e_1-\frac{1}{4}e_2-\frac{1}{4}e_3-\frac{1}{4}e_4,{}\\
\lambda_2&=\frac{1}{2}\alpha_1+\alpha_2+\frac{1}{2}\alpha_3=
\frac{1}{2}e_1+\frac{1}{2}e_2-\frac{1}{2}e_3-\frac{1}{2}e_4,{}\\
\lambda_3&=\frac{1}{4}\alpha_1+\frac{1}{2}\alpha_2+\frac{3}{4}\alpha_3
=\frac{1}{4}e_1+\frac{1}{4}e_2+\frac{1}{4}e_3-\frac{3}{4}e_4.
\end{split}
\end{equation}
Weyl vector has the form

\begin{equation}\label{3-4}
\rho = \frac{1}{2}(3\alpha_1+4\alpha_2+3\alpha_3)=\lambda_1+\lambda_2+\lambda_3=
\frac{1}{2}(3e_1+e_2-e_3-3e_4).
\end{equation}

The Weyl group $W(A_3)$ is identified with the group of permutations of the vectors 
$e_i,\,i=1,..,4,$ with $4!=24$ elements. The generators of this group are the elements 
$w_1,w_2,w_3$ which permutate the vectors $e_1$  and $e_2$, $e_2$ and $e_3$, $e_3$ 
and $e_4$ respectively.

{\bf 4.}
To begin with, let us calculate the orbit function $\Phi_{\bf n}({\boldsymbol{\phi}})$
(\ref{0-7}) associated with the Lie algebra $A_3$.  Following the scheme presented above,  
calculate the scalar product $(w\lambda_k,{\boldsymbol{\phi}})$ for all the elements 
$w\in W(A_3)$, taking into account the notation
${\boldsymbol{\phi}} = \sum_i\phi_i\alpha_i$ and the representation  (\ref{dop-4}) 
of the roots and fundamental weights by the orthogonal vectors  $e_i$. The result 
can be written in the form of diagonal matrices

\begin{equation}
\begin{split}
A_1 = {\rm diag}&(\phi_1,-\phi_1+\phi_2,\phi_1,\phi_1,-\phi_1+\phi_2,-\phi_2+\phi_3,
-\phi_1+\phi_2,\phi_1, \phi_1,-\phi_2+\phi_3,-\phi_1+\phi_2,
{}\nonumber \\
&-\phi_3,\phi_1,-\phi_1+\phi_2,-\phi_2+\phi_3,-\phi_2+\phi_3,-\phi_3,-\phi_3,
-\phi_1+\phi_2,-\phi_2+\phi_3,-\phi_3,-\phi_3,
 {} \nonumber \\
&-\phi_3,-\phi_2+\phi_3)  {} \nonumber \\
A_2 = {\rm diag}&(\phi_2,\phi_2,\phi_1-\phi_2+\phi_3,\phi_2,-\phi_1+\phi_3,\phi_1-\phi_2+\phi_3,
\phi_2,\phi_1-\phi_2+\phi_3,\phi_1-\phi_3, {}\nonumber \\
&-\phi_1+\phi_3,\phi_1-\phi_3,-\phi_1+\phi_3,\phi_1-\phi_2+\phi_3,
-\phi_1+\phi_2-\phi_3,\phi_1-\phi_3,-\phi_1+\phi_3, {}\nonumber \\
&-\phi_1+\phi_2-\phi_3,\phi_1-\phi_3,-\phi_1+\phi_2-\phi_3,-\phi_2,-\phi_1+\phi_2
-\phi_3,-\phi_2,-\phi_2,-\phi_2), {} \nonumber \\
A_3 = {\rm diag}&(\phi_3,\phi_3,\phi_3,\phi_2-\phi_3,\phi_3,\phi_3,\phi_2-\phi_3,\phi_1-\phi_2,
\phi_2-\phi_3,\phi_3,\phi_1-\phi_2,-\phi_1,\phi_1-\phi_2, {}  \nonumber\\
&\phi_2-\phi_3,\phi_2-\phi_3,-\phi_1,\phi_2-\phi_3,
\phi_1-\phi_2,-\phi_1,-\phi_1,-\phi_1,\phi_1-\phi_2,-\phi_1,\phi_1-\phi_2).{} \nonumber \\
\end{split}
\end{equation}
Using these matrices one can express the function $\Phi_{\bf n}({\boldsymbol{\phi}})$
in the form (\ref{dop-6})

\begin{equation}
\label{1-4}
\Phi_{\bf n}({\boldsymbol{\phi}}) = {\rm tr}(e^{2\pi\ri(n_1A_1+n_2A_2+n_3A_3)}).
\end{equation}
Denoting by $M_k$ the exponential $e^{2\pi\ri A_k}$, let us introduce the matrices  
$R_k=(I_{24} - p_kM_k)^{-1},\,k=1,2,3,$  where $p_k$  are real parameters, $I_{24}$ 
is the identity $24 \times 24$ matrix. In the terms of $R_k$ the orbit function 
$\Phi_{\bf n}({\boldsymbol{\phi}})$ has the form

\begin{equation}
\label{1-5}
\Phi_{{n_1,n_2,n_3}}({\boldsymbol{\phi}}) =
\frac1{n_1!n_2!n_3!}\left.\frac{\rd^{(n_1+n_2+n_3)}}{\rd^{n_1}p_1\rd^{n_2}p_2\rd^{n_3}p_3}
\left({\rm tr}(R_{p_1}R_{p_2}R_{p_3})
\right)\right|_{p_1= p_2 =p_3 = 0}.
\end{equation}
From (\ref{1-5}) it follows that the function 
$F^I_{{p_1,p_2,p_3}}={\rm tr}(R_{p_1}R_{p_2}R_{p_3})$ 
expressed in terms of new variables is the generating function of three variables Chebyshev 
polynomials of the first kind.
Calculation of the function (\ref{1-5}) for the indices  $(1,0,0),(0,1,0),(0,0,1)$ gives us

$$\Phi_{{1,0,0}} = 6z,\quad \Phi_{{0,1,0}}=4r,\quad \Phi_{{0,0,1}}=6\bar z,$$
where the new variables  are defined  by

\begin{equation}
\begin{split}\label{dop-9}
z &= e^{2\pi\ri\phi_1}+e^{2\pi\ri(-\phi_1+\phi_2)}+e^{2\pi\ri(-\phi_2+\phi_3)}
+e^{-2\pi\ri\phi_3},{} \\
r &= e^{2\pi\ri\phi_2}+e^{2\pi\ri(\phi_1-\phi_2+\phi_3)}+e^{2\pi\ri(-\phi_1+\phi_3)}
+e^{2\pi\ri(\phi_1-\phi_3)}+e^{2\pi\ri(-\phi_1+\phi_2-\phi_3)}+e^{-2\pi\ri\phi_2},{} \\
\bar z &= e^{2\pi\ri\phi_3}+e^{2\pi\ri(-\phi_3+\phi_2)}+e^{2\pi\ri(-\phi_2+\phi_1)}
+e^{-2\pi\ri\phi_1}.{}
\end{split}
\end{equation}
In the terms of $z,\bar z,r$  generating function $F^I_{{p_1,p_2,p_3}}$ is presented in the following form

\begin{equation}
\label{1-9}
F^I_{{p_1,p_2,p_3}}=\frac{\sum\limits _{i,j,k=0}^{i,k=3;j=5}K_{i,j,k}(z,r,\bar z)p_1^ip_2^jp_3^k}{Z_{p_1}Z_{p_2}Z_{p_3}},
\end{equation}
where

\begin{equation}
\begin{split}\label{dop-10}
Z_{p_1}&=1-zp_1+rp_1^2-\bar z p_1^3+p_1^4,{} \\
Z_{p_2}&=1-rp_2+(z\bar z-1)p_2^2-(z^2+\bar z^2-2r)p_2^3+(z\bar z-1)p_2^4-rp_2^5+p_2^6,{} \\
Z_{p_3}&=1-\bar z p_3+rp_3^2-zp_3^3+p_3^4.{}
\end{split}
\end{equation}
It is not difficult to calculate the coefficients $K_{i,j,k}(z,r,\bar z)$ from (\ref{1-9}). 
In the Appendix A only those coefficients $K_{i,j,k}$ are listed whose subscripts satisfy the 
inequality $i\le k$. This is because of the symmetry condition 
${\overline K}_{i,j,k}(z,r,\bar z)=K_{k,j,i}(\bar z,r,z)$.

Calculation of the orbit function $\Phi_{{n_1,n_2,n_3}}$ with the subscripts $(1,0,0),(0,1,0)$ 
using the generating function (\ref{1-9}) gives us $\Phi_{{1,0,0}} = 6z,\, \Phi_{{0,1,0}}=4r$.
The reason for appearance of the integer factors in these formulas is the presence of stabilizers 
of weight lattice vectors. By definition, stabilizer of $\bf n$ is a subgroup of a Weyl group $W$ 
consisting of all elements $w\in W$, such that $w{\bf n} = {\bf n}$. In the work \cite{KP} were 
obtained the stabilizers in the Weyl group $W(A_3)$ of all vectors $\bf n$ from the weight lattice. 
In accordance with this result it is convenient the following normalization of the three variables 
Chebyshev polynomials $T_{{n_1,n_2,n_3}}$ of the first kind

$$
T_{{0,0,0}}=\frac{\Phi_{{0,0,0}}}{24},\quad T_{{n_1,0,0}}=\frac{\Phi_{{n_1,0,0}}}{6},
\quad T_{{0,n_2,0}}=\frac{\Phi_{{0,n_2,0}}}{4},\quad T_{{0,0,n_3}}=\frac{\Phi_{{0,0,n_3}}}{6}$$
$$T_{{n_1,n_2,0}}=\frac{\Phi_{{n_1,n-2,0}}}{2},\quad T_{{0,n_2,n_3}}=\frac{\Phi_{{0,n_2,n_3}}}{2},\quad
\quad T_{{n_1,0,n_3}}=\frac{\Phi_{{n_1,0,n_3}}}{2},\quad T_{{n_1,n_2,n_3}}=\Phi_{{n_1,n_2,n_3}}.
$$
In this normalization all polynomial coefficients are integer. In the Appendix B are listed first 
three variables Chebyshev polynomials of the first kind associated with the Lie algebra $A_3$.

{\bf 5.} In this section we calculate the generating function of three variables Chebyshev 
polynomials of the second kind. The function (\ref{0-9}) can be represented in the form

$$
U_{n_1,n_2,n_3}=\frac{\Phi_{n_1+1,n_2+1,n_3+1}^{as}}{\Phi_{1,1,1}^{as}},
$$
because of the Weyl vector is the sum of the fundamental weights 
${\bf\rho} = \lambda_1+\lambda_2+\lambda_3$ (\ref{3-4}).
Let us rewrite the function $\Phi_{\bf n}^{as}$ as the difference 
$\Phi_{\bf n}^{as+}-\Phi_{\bf n}^{as-}$ according to sign of $\det{w}$ and introduce 
the following diagonal matrices

\begin{equation}
\begin{split}
A_{1+}&={\rm diag}(\phi_1,-\phi_1+\phi_2,-\phi_2+\phi_3,-\phi_1+\phi_2,\phi_1,\phi_1,-\phi_2+\phi_3,
-\phi_3,-\phi_3,-\phi_1+\phi_2,{}\nonumber \\
&-\phi_3,-\phi_2+\phi_3),{}\nonumber \\
A_{2+}&={\rm diag}(\phi_2,-\phi_1+\phi_3,\phi_1-\phi_2+\phi_3,\phi_2,\phi_1-\phi_2+\phi_3,\phi_1-\phi_3,
-\phi_1+\phi_3,-\phi_1+\phi_2-\phi_3,{}\nonumber \\
&\phi_1-\phi_3,-\phi_1+\phi_2-\phi_3,-\phi_2,-\phi_2),{} \\
\end{split}
\end{equation}

\begin{equation}
\begin{split}
A_{3+}&={\rm diag}(\phi_3,\phi_3,\phi_3,\phi_2-\phi_3,\phi_1-\phi_2,\phi_2-\phi_3,-\phi_1,\phi_2-\phi_3,
\phi_1-\phi_2,-\phi_1,-\phi_1,\phi_1-\phi_2),{}\nonumber \\
A_{1-}&={\rm diag}(-\phi_1+\phi_2,\phi_1,\phi_1,-\phi_2+\phi_3,\phi_1,-\phi_1+\phi_2,-\phi_2+\phi_3,
-\phi_1+\phi_2,-\phi_3,-\phi_2+\phi_3,{}\nonumber \\
&-\phi_3,-\phi_3),{}\nonumber \\
A_{2-}&={\rm diag}(\phi_2,\phi_1-\phi_2+\phi_3,\phi_2,-\phi_1+\phi_3,\phi_1-\phi_3,-\phi_1+\phi_3,
\phi_1-\phi_2+\phi_3,-\phi_1+\phi_2-\phi_3,{}\nonumber \\
&\phi_1-\phi_3,-\phi_2,-\phi_1+\phi_2-\phi_3,
-\phi_2),{}\nonumber \\
A_{3-}&=  {\rm diag}(\phi_3,\phi_3,\phi_2-\phi_3,\phi_3,\phi_1-\phi_2,-\phi_1,\phi_1-\phi_2,\phi_2-\phi_3,
\phi_2-\phi_3,-\phi_1,-\phi_1,\phi_1-\phi_2).{}\nonumber
\end{split}
\end{equation}
In the terms of these matrices the function from nominator of (\ref{0-9}) has the form

$$
\Phi_{n_1+1,n_2+1,n_3+1}^{as}={\rm tr}(M_{1+}^{n_1+1}M_{2+}^{n_2+1}M_{3+}^{n_3+1}-
M_{1-}^{n_1+1}M_{2-}^{n_2+1}M_{3-}^{n_3+1}),
$$
where $M_{k\pm}=exp{(2\pi\ri A_{k\pm})}$. Define the following  matrices
$R_{k\pm}=(I_{12} - p_kM_{k\pm})^{-1},\,k=1,2,3,$  where $p_k$  are real 
parameters and $I_{12}$ is the identity $12 \times 12$ matrix.  In the terms of 
$M_{k\pm}$ and $R_{k\pm}$ matrices the generating function of three variables 
Chebyshev polynomials of the second kind has the form

$$
F^{II}_{{p_1,p_2,p_3}}=\frac{{\rm tr}(R_{1+}R_{2+}R_{3+}-R_{1-}R_{2-}R_{3-})}
{{\rm tr}(M_{1+}M_{2+}M_{3+}-M_{1-}M_{2-}M_{3-})}.
$$
Using the variables $z,r,\bar z$ introduced previously (\ref{dop-9}) 
we finally obtain the following function

$$F^{II}_{{p_1,p_2,p_3}}=\frac{L_1(p_1,p_2,p_3)+zL_2(p_1,p_2,p_3)+rL_3(p_1,p_2,p_3)+
\bar zL_4(p_1,p_2,p_3)}{Z_{p_1}Z_{p_2}Z_{p_3}}$$
where
\begin{equation}
\begin{split}
L_1(p_1,p_2,p_3)&=1-p_2^2-p_1p_3-p_1^2p_2^4p_3^2-p_1^2p_2^3-p_2^3p_3^2+
p_1^2p_2^2p_3^2+p_1^2p_2+p_2p_3^2+p_1p_2^4p_3,{}\nonumber \\
L_2(p_1,p_2,p_3)&=p_3^2p_2^3p_1+p_1p_2^2-p_2p_3-p_1^2p_2^2p_3,{}\nonumber \\
L_3(p_1,p_2,p_3)&=-p_3p_2^3p_1+p_2p_3p_1,{}\nonumber \\
L_4(p_1,p_2,p_3)&=p_3p_2^2+p_3p_2^3p_1^2-p_1p_2^2p_3^2-p_1p_2,{}\nonumber
\end{split}
\end{equation}
and $Z_{p_i}$ were defined by (\ref{dop-10}). Some first polinomials calculated  
via this function are listed in Appendix C.

\vspace{1cm}

\noindent{\bf Acknowledgments}

\noindent The author are thankful to professor E.V. Damaskinsky for discussions and comments. 
This work has been supported by RFFR grant no. 15-01-03148-а.
\vspace{5mm}

\newpage

{\large\bf Appendix A} The list of coefficients $K_{i,j,k}(z,r,\bar z)$ 
for the generating function of
three variables Chebyshev polynomials of the first kind 
(${\overline K}_{i,j,k}(z,r,\bar z)=K_{k,j,i}(\bar z,r,z)$)

\begin{align}
K_{000}&=24, & K_{030}&=24r-12z^2-12\bar z^2,\nonumber\\
K_{001}&=-18\bar z, & K_{031}&=10z^2\bar z-28r\bar z-6z+12\bar z^3,\nonumber\\
K_{002}&=12r, & K_{032}&=-8z^2r-83\bar z^2r+16r^2+16z\bar z-16,\nonumber\\
K_{003}&=-6z, & K_{033}&=6z^3-16rz+4\bar z^2z-4\bar z,\nonumber\\
K_{010}&=-20r, & K_{040}&=8z\bar z-8,\nonumber\\
K_{011}&=16r\bar z-6z, & K_{041}&=4rz+14\bar z-8\bar z^2z,\nonumber\\
K_{012}&=-12r^2+4z\bar z+8, & K_{042}&=-6\bar z^2-6z^2-4r+6zr\bar z,\nonumber\\
K_{013}&=-6\bar z+6rz,& K_{043}&=10z+4\bar zr-4z^2\bar z,\nonumber\\
K_{020}&=16z\bar z-16, & K_{050}&=-4r,\nonumber\\
K_{021}&=20 \bar z+4 rz-14 z\bar z^2,&K_{051}&=-6z+4r\bar z,\nonumber\\
K_{022}&= 10rz\bar z-6z^2-6\bar z^2-8r,&K_{052}&=-4r^2+4z\bar z-8,\nonumber\\
K_{023}&=-6\bar zz^2+4r\bar z+12z, & K_{053}&=12zr-6\bar z,\nonumber
\end{align}
\begin{align}
K_{101}&=14z\bar z-8, & K_{121}&=12z^2\bar z^2-2\bar z^2r-2z^2r-4r^2-24z\bar z,\nonumber\\
K_{102}&=-10zr+6\bar z, &K_{122}&=-4\bar z+3z^3+16zr-9z^2\bar zr+7z\bar z^2+2\bar zr^2,\nonumber\\
K_{103}&=6z^2-4r, & K_{123}&=6z^3\bar z-8r-7rz\bar z+3\bar z^2-9z^2,\nonumber\\
K_{111}&=12r+3z^2+3\bar z^2-13z\bar zr, &K_{131}&=-10\bar z^3z-10z^3\bar z+7\bar z^2+7z^2-8r+29rz\bar z,\nonumber\\
K_{112}&=10r^2z-10z-2z^2\bar z-8r\bar z, &K_{132}&=8z^3r+12z-3\bar z^3+7z\bar z^2r-15z^2\bar z-18r^2z+4\bar zr,\nonumber\\
K_{113}&=4r^2-6z^2r+4z\bar z+8,& K_{133}&=-6z^4-4\bar z^2z^2+2\bar z^2r+18z^2r+4z\bar z-4r^2,\nonumber
\end{align}
\begin{align}
K_{141}&=8z^2\bar z^2-11z\bar z-4z^2r-4\bar z^2r+4r^2+8, & K_{151}&=-4r+6z^2+6\bar z^2-4rz\bar z,\nonumber\\
K_{142}&=-2\bar z+6z^3-6z^2\bar zr+2\bar zr^2+8z\bar z^2+2zr,& K_{152}&=4r^2z-2z-4z^2\bar z-4r\bar z,\nonumber\\
K_{143}&=4z^3\bar z-5z\bar zr-13z^2-3\bar z^2+12r,&K_{153}&=-2z^2r+8z\bar z-8,\nonumber
\end{align}
\begin{align}
K_{202}&=8r^2-4z\bar z-8,& K_{212}&=-8r^3+5rz\bar z+3z^2+3\bar z^2+12r,\nonumber\\
K_{203}&=-4zr+6\bar z,&K_{213}&=4zr^2-8r\bar z-6z,\nonumber
\end{align}
\begin{eqnarray}
K_{222}&=&7r^2z\bar z-2z^2\bar z^2-5rz^2-5r\bar z^2-8r^2-8z\bar z+16,\nonumber\\
K_{223}&=&-4\bar z-4rz^2\bar z+4z\bar z^2+12zr+2r^2\bar z,\nonumber\\
K_{232}&=&-6r^2z^2-6r^2\bar z^2-24r+4z^2+4\bar z^2+2\bar z^3z+2z^3\bar z+12r^3+12rz\bar z,\nonumber\\
K_{233}&=&3\bar z^2zr-10r^2z-7\bar zz^2+4z^3r-3\bar z^3+4r\bar z+4z,\nonumber\\
K_{242}&=&24z\bar zr^2-2z^2\bar z^2-7z^2r-7\bar z^2r+4z\bar z-8,\nonumber\\
K_{243}&=&-3rz^2\bar z+5z\bar z^2+3z^3+2r^2\bar z-2\bar z,\nonumber
\end{eqnarray}

\begin{align}
K_{252}&=-4r^3+12r+7rz\bar z-3z^2-3\bar z^2,&K_{323}&=2z^2\bar z^2-8z\bar z-4r^2+8z\bar z,\nonumber\\
K_{253}&=2r^2z-2\bar zz^2-4r\bar z+2z,&K_{333}&=-2\bar zz^3-2z\bar z^3+7rz\bar z+5z^2+5\bar z^2-8r,\nonumber\\
K_{303}&=2z\bar z-8,& K_{343}&=2z^2\bar z^2-2r\bar z^2-2rz^2+4r^2-10z\bar z+8,\nonumber\\
K_{313}&=-2rz\bar z+12r,& K_{353}&=-1rz\bar z+3z^2+3\bar z^2-4r.\nonumber
\end{align}

{\large\bf Appendix B} The list of first three variables Chebyshev polynomials of the first kind (${\overline T}_{i,j,k}(z,r,\bar z)=T_{k,j,i}(\bar z,r,z)$)

\begin{align}
T_{000}&=1, & T_{111}&=rz{\bar z}-3{\bar z}^2+4r-3z^2,\nonumber\\
T_{100}&=z, & T_{030}&=-3rz{\bar z}-3r+3z^2+3{\bar z}^2+r^3,\nonumber\\
T_{010}&=r, & T_{400}&=4z{\bar z}-4z^2r+z^4+2r^2-4,\nonumber\\
T_{200}&=z^2-2r, & T_{310}&=5{\bar z}r-3r^2z+z^3r-{\bar z}z^2+z,\nonumber\\
T_{110}&=rz-3{\bar z},& T_{301}&=3{\bar z}^2-3rz{\bar z}+z^3{\bar z}+2r-z^2,\nonumber\\
T_{101}&=z{\bar z}-4,& T_{220}&=2r+4rz{\bar z}-2r^3+2z^2-2z^3{\bar z}+r^2z^2-3{\bar z}^2,\nonumber\\
T_{020}&=2-2z{\bar z}+r^2, &T_{211}&=-2r^2{\bar z}+z^2{\bar z}r-z{\bar z}^2-2{\bar z}+8rz-3z^3,\nonumber\\
T_{300}&=3{\bar z}-3rz+z^3, &T_{202}&=-2{\bar z}^2r+z^2{\bar z}^2+4r^2-2z^2r-4,\nonumber\\
T_{210}&=-2r^2+z^2r-z{\bar z}+4, &T_{130}&=-2rz-r^2{\bar z}-3z^2{\bar z}r+3z^3+5z{\bar z}^2+zr^3-5{\bar z},\nonumber\\
T_{201}&=-2{\bar z}r+{\bar z}z^2-z, &T_{121}&=10z{\bar z}-2z^2{\bar z}^2+r^2z{\bar z}-{\bar z}^2r-z^2r-8,\nonumber\\
T_{120}&=5z-2{\bar z}z^2+r^2z-{\bar z}r,&T_{040}&=6-8z{\bar z}+4{\bar z}^2r-4r^2+4z^2r+3z^2{\bar z}^2-4r^2z{\bar z}+r^4.\nonumber
\end{align}

{\large\bf Appendix C} The list of first three variables Chebyshev polynomials of the second kind (${\overline U}_{i,j,k}(z,r,\bar z)=U_{k,j,i}(\bar z,r,z)$)
\begin{align}
U_{000} &= 1, &U_{111} &= rz{\bar z}-{\bar z}^2-z^2,\nonumber\\
U_{100} &= z, &U_{030} &= -r-2rz{\bar z}+z^2+{\bar z}^2+r^3,\nonumber\\
U_{010} &= r, &U_{400} &= -1+2z{\bar z}+r^2-3rz^2+z^4,\nonumber\\
U_{200} &= -r+z^2, &U_{310} &= 2{\bar z}r-2zr^2+rz^3-{\bar z}z^2+z,\nonumber\\
U_{110} &= zr-{\bar z}, &U_{301} &= {\bar z}^2-2rz{\bar z}+z^3{\bar z}+r-z^2,\nonumber\\
U_{101} &= -1+z{\bar z}, &U_{220} &= r-r^3+z^2-z^3{\bar z}+r^2z^2,\nonumber\\
U_{020} &= -z{\bar z}+r^2, &U_{211} &= {\bar z}-{\bar z}r^2+{\bar z}rz^2-z{\bar z}^2+zr-z^3,\nonumber\\
U_{300} &= {\bar z}-2zr+z^3, &U_{202} &= -z{\bar z}+r^2-r{\bar z}^2-rz^2+z^2{\bar z}^2,\nonumber\\
U_{210} &= -r^2+rz^2+1-z{\bar z}, &U_{130} &= -2{\bar z}rz^2+z^3+2z{\bar z}^2+zr^3-{\bar z}-{\bar z}r^2,\nonumber\\
U_{201} &= -{\bar z}r+{\bar z}z^2-z, &U_{121} &= -r{\bar z}^2+3z{\bar z}-z^2{\bar z}^2+r^2z{\bar z}-1-rz^2,\nonumber\\
U_{120} &= z-{\bar z}z^2+zr^2-{\bar z}r, &U_{040} &= 2rz^2+2r{\bar z}^2-2r^2+1-2z{\bar z}+z^2{\bar z}^2-3r^2z{\bar z}+r^4.\nonumber
\end{align}


\begin{thebibliography}{99}
\bibitem{K1}
T.N. Koornwinder, {\it Orthogonal polynomials in two variables which are eigenfunctions of two algebraically independent partial differential operators I-IV}. Indagationes Mathematicae Proc. {\bf 77}, 48-66, 357-81 (1974).
\bibitem{H}
G.J. Heckman, {\it Root systems and hypergeometric functions II}. Comp. Math. {\bf 64}, 353-73, (1987).
\bibitem{HW}
M.E. Hoffman, W.D. Withers, {\it Generalized Chebyshev polynomials associated with affine Weyl groups}. Trans. Am. Math. Soc., {\bf 308}, 91-104 (1988).
\bibitem{B}
R.J. Beerends, {\it Chebyshev polynomials in several variables and the radial part Laplace-Beltrami operator}. Trans. Am. Math. Soc., {\bf 328}, 770-814, (1991).
\bibitem{KP}
A. Klimyk, J.Patera, {\it Orbit functions}. SIGMA. {\bf 2} 006, (2006).
\bibitem{LU} V.D. Lyakhovsky, Ph.V. Uvarov, {\it Multivariate Chebyshev polynomials}. J. Phys. A: Math. Theor. {\bf 46}, 125201, (2013).
\bibitem{RM-K}
B.N. Ryland, H.Z. Munthe-Kaas, {\it On multivariate Chebyshev polynomials and spectral approximations on triangles Spectral and High OrderMethods for Partial Differential Equations}. Lecture Notes in Computer Science and Engineering, {\bf 76}, Berlin, Springer, 19 - 41, (2011).
\bibitem{SS} B. Shapiro, M. Shapiro, {\it On Eigenvalues of Rectangular Matrices}. Тр. МИАН, {\bf 267}, 258-265, (2009).
\bibitem{KLP} P.P. Kulish, V.D. Lyakhovsky, O.V. Postnova, {\it Multiplicity function for tensor powers of modules of the $A_n$ algebra}. Theoretical and Mathematical Physics, {\bf 171}, 283-293, (2012) (in Russian).
\bibitem{KLP2} P.P. Kulish, V.D. Lyakhovsky, O.V. Postnova, {\it Tensor power decomposition. $B_n$-case}.
Journal of Physics: Conference Series, {\bf 343}, 012095, (2012).
\bibitem{KLP3} P.P. Kulish, V.D. Lyakhovsky, O.V. Postnova, {\it Tensor powers for non-simply laced Lie algebras
$B_2$-case}. Journal of Physics: Conference Series {\bf 346} 012012 (2012).
\bibitem{L} V.D. Lyakhovsky, {\it Multivariate Chebyshev polynomials in terms of singular elements}. Theoretical and Mathematical Physics, {\bf 175}, 797-805, (2013).
\bibitem{BD} V.V. Borzov, E.V. Damaskinsky, {\it Chebyshev - Koornwinder oscillator}. Theoretical and Mathematical Physics, {\bf 175}, 765-772, (2013).
\bibitem{BD1} V.V. Borzov, E.V. Damaskinsky, {\it The algebra of two dimensional generalized Chebyshev-Koornwinder oscillator}. Journal of Mathematical Physics, {\bf 55}, 103505, (2014).
\bibitem{GR} G. Von Gehlen, S. Roan, {\it The superintegrable chiral Potts quantum chain and generalized Chebyshev polynomials}. Integrable Structures of Exactly Solvable Two- Dimensional Models of Quantum Field Theory, eds. S.
     Pakuliak, G. Von Gehlen, NATO Science Series, {\bf 35}, 155-172, Berlin, Springer, (2001).
\bibitem{G}  Von Gehlen G 2002 {\it Onsager's algebra and partially orthogonal polynomials}. Int. J. Mod. Phys., {\bf B 16} 2129
\bibitem{S} P.K. Suetin, {\it Classical orthogonal polynomials}. Moskva, Nauka (1979) (in Russian).
\bibitem{NBurb} Burbaki N 1975 Elements de Mathematique. Groupes et Algebres de Lie (Paris: Hermann).
\bibitem{DL}Ken B. Dunn, R. Lidl, {\it Generalizations of the classical Chebyshev polynomials to polynomials in two variables}. Czech. Math. J., {\bf 32}, 516-528, (1982).
\bibitem{DKS} E.V. Damaskinsky, P.P. Kulish, M.A. Sokolov, {\it On calculation of generating functions of multivariate Chebyshev polynomials}. POMI preprint 13/2014  (in Russian).
\bibitem{SJ} SUN JiaChang,  {\it A new class of three-variable orthogonal polynomials and their recurrences relations}. Science in China, Series A: Mathematics, Jun., {\bf  51}, 1071-1092, (2008).
\end{thebibliography}
\end{document}